\documentclass[fleqn,twoside]{article}
\usepackage{espcrc2}
\usepackage{epsfig}
\usepackage{color}

\usepackage{graphicx}
\usepackage[figuresright]{rotating}

\newcommand{\AmS}{{\protect\the\textfont2
  A\kern-.1667em\lower.5ex\hbox{M}\kern-.125emS}}
\def\xlf{\raisebox{+0.2em}{\color{red}\boldmath{$\chi$}}\hspace{-0.2ex}\raisebox{-0.2em}{\color{green}L}
\hspace{-1.5ex}\raisebox{+0.14em}{\color{blue}F}\hspace{2mm}}                   
\input macros.sty      

\title{
Comparison between overlap and twisted mass fermions towards the chiral limit\thanks{Talk presented by M.~Papinutto.}}  

\author{\xlf
    Coll.:~W.~Bietenholz\address[HUMB]{\vspace*{-0.2cm}Humboldt
    Universit\"{a}t zu Berlin, Institut f\"{u}r Physik, Newtonstr.\ 15,
    D-12489 Berlin, Germany},
    S.~Capitani\address[DESY]{\vspace*{-0.2cm}NIC/DESY
    Zeuthen, Platanenallee 6, D-15738 Zeuthen,
    Germany}\address[GRAZ]{\vspace*{-0.2cm}Institut f\"ur Physik/Theoretische Physik,
    Universit\"at Graz, A-8010, Austria}, 
    T.~Chiarappa\addressmark[DESY], 
    M.~Hasenbusch\address[SOTON]{\vspace*{-0.2cm}School of Physics and Astronomy, University
    of Southampton, Southampton, SO17 1BJ, UK},
    K.~Jansen\addressmark[DESY],
    K.-I.~Nagai\addressmark[DESY],
    M.~Papinutto\addressmark[DESY],
    L.~Scorzato\addressmark[HUMB],
    S.~Shcheredin\addressmark[HUMB],
    A.~Shindler\addressmark[DESY],
    C.~Urbach\addressmark[DESY]\address{\vspace*{-0.0cm}Freie
    Universit\"{a}t Berlin, Institut f\"{u}r Theoretische Physik,
    Arnimallee 14, D-14195 Berlin, Germany}, U.~Wenger
    \addressmark[DESY], I.~Wetzorke\addressmark[DESY].}

\begin{document}

\begin{abstract}
We compare overlap fermions, which are chirally 
invariant, and Wilson twisted mass fermions in the approach 
to the chiral limit. Our quenched simulations reveal that 
with both formulations of lattice fermions pion masses of O(250 MeV) 
can be reached in practical simulations.
Our comparison is done at a fixed lattice spacing $a\simeq
0.123$ fm. Several quantities are measured, such as hadron masses and
pseudoscalar decay constants.
\vspace{-0.22cm}
\end{abstract}

\maketitle

\section{INTRODUCTION}
\vspace*{-0.2cm}
The Wilson formulation of lattice QCD exhibits various problems: 
the presence of unphysical small eigenvalues which give rise to 
exceptional configurations, the explicit breaking
of chiral symmetry which complicates the pattern of operator mixing,
the presence of large discretization errors which are reduced through 
the Symanzik improvement program. In the present study we consider two
formulations of lattice QCD that are able to overcome most of these
problems: overlap
and twisted mass (tm) fermions.
Overlap fermions have an exact chiral symmetry at finite lattice 
spacing $a$
 and the mass is an infrared cut-off, thus allowing 
the approach to the chiral limit to be performed at finite $a$. 
For tm fermions the twisted part of the mass also provides 
an infrared cut-off, thus solving the
practical problem of exceptional configurations which affects Wilson
fermions. At maximal twist angle, moreover, one has
automatic O($a$) improvement for various quantities like energy 
eigenvalues and matrix elements~\cite{Frezzotti:2003ni}.  
The great advantages of overlap fermions have unfortunately
the price of being rather expensive from the numerical point of view. 
On the other hand, tm fermions are rather cheap to simulate but show
residual chiral symmetry breaking effects.

Our comparison is performed at a fixed value of $\beta=5.85$
($a^{-1}\simeq 1.605$ GeV) and no attempt of a scaling 
analysis~\cite{JSUW} is performed here. Our aim is to investigate how both 
formulations behave in their approach to the chiral limit for a number 
of quantities like hadron masses and pseudoscalar decay 
constants. We also 
provide a timing estimate, from the results in ref.~\cite{solver}.

\vspace*{-0.3cm}
\section{NUMERICAL RESULTS}
\vspace*{-0.2cm}
For overlap fermions we have 140 configurations on $12^3\times 24$ 
lattices ($L_{12}\sim 1.48$ fm). The bare quark
masses are $m_\mathrm{ov} a=0.01,0.02,0.04,0.06,0.08,0.10$ and
$\rho=1.6$~\cite{solver}. 
The simulations for tm fermions are done at 
full twist so that all the quantities studied here ought to be automatically
O($a$) improved~\cite{Frezzotti:2003ni}. In the following we refer to 
the ``twisted'' basis where the action is the normal Wilson action with a 
bare mass term $M_{\rm cr}+i\mu\gamma_5\tau_3$.
$M_{\rm cr}$ is the bare critical mass determined for normal
Wilson fermions and, for the corresponding value of 
the hopping parameter, we have used $\kappa_{\rm cr}=0.16166(2)$~\cite{JSUW}.
The twisted quark mass parameter $\mu a$ has been chosen to have the 
same values of the overlap masses plus the value 0.005 
(in the plot and tables below both 
$m_\mathrm{ov}$ and $\mu$ will be called $m_{{\rm bare}}$). 
We collected 140 configurations on $12^3\times 24$,
140 configurations on $14^3\times 32$ ($L_{14}\sim 1.72$ fm) and
250 configurations on $16^3\times 32$ ($L_{16}\sim 1.97$ fm) lattices.
For both, overlap and tm fermions, a multiple mass solver (MMS) has been
employed.  

\vspace{-0.0cm}
\subsection{HADRON MASSES}

\begin{table}[b]
\vspace*{-0.9cm}
\hspace*{-0.4cm}
{\footnotesize\begin{tabular}{ccccc}
\hline
\hline
$m_{{\rm bare}} a$&
$M^{P-S}_{\pi,{\rm ov}}a$ &
$M^{P}_{\pi,{\rm tm}}a$
& $f^{{\rm ov}}_{\pi}a$ & $f^{{\rm tm}}_{\pi}a$ \\
\hline
0.005 & - & 0.1694(24)& - & 0.0794(17)\\
0.01 & 0.140(20) & 0.2276(22)& 0.0934(90)&0.0904(12)\\
0.02 & 0.196(14) & 0.3141(20)& 0.1012(53)&0.1022(12)\\
0.04 & 0.280(10) & 0.4468(18)& 0.1060(34)&0.1170(13)\\
0.06 & 0.346(8)  & 0.5552(15)& 0.1106(25)&0.1295(12)\\
0.08 & 0.401(7)  & 0.6505(13)& 0.1157(22)&0.1411(12)\\
0.10 & 0.451(6)  & 0.7373(14)& 0.1209(21)&0.1522(13)\\
\hline
\hline
\end{tabular}}
\vspace*{0.0cm}
\caption{\mbox{$M_\pi a$ and $f_\pi a$ for overlap and tm fermions.}
\label{tab:mmesons1}}
\vspace*{-0.3cm}
\end{table}

\begin{figure}[t]
\vspace{0.1cm}
\hspace*{-0.5cm}
\epsfig{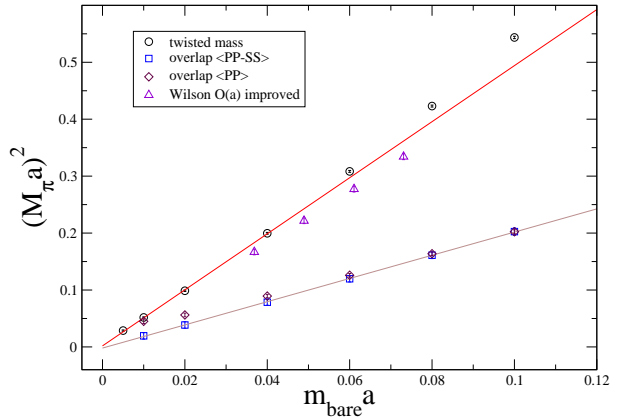}
\vspace{-1.4cm}
\caption{ \label{fig:mpi_ov_tm}
$M_\pi^2 a^2$ vs. $m_{{\rm bare}}a$.}
\vspace{-0.9cm}
\end{figure}

We extract hadron masses by fitting the behaviour of suitable two point
functions at large euclidean time\footnote{In the overlap case the
spectrum is automatically O($a$) improved. Moreover, in order to obtain
O($a$) improved estimates of the the decay constants the 
bilinears can also be easily improved~\cite{giusti}.}. We check in various ways 
(including the use of Jacobi smearing) that our determination is 
not contaminated by the presence of excited states.
Pseudoscalar masses are extracted from the correlation functions
$C_{P}^b(x_0) = \sum_{\mathbf x} \langle P^b(x)P^b(0)\rangle$ 
and (only for overlap) $C_{P-S}(x_0) = \sum_{\mathbf x} \langle
P^b(x)P^b(0)-S^b(x)S^b(0)\rangle$
where $b$ is the flavour index and, in order to avoid problems of mixing
with the scalar density, in the tm case we only consider $a=1,2$.
In $C_{P-S}(x_0)$ the contribution of the topological zero modes
of the overlap operator cancels. 
Results are reported in 
Fig.~\ref{fig:mpi_ov_tm} and in Tab.~\ref{tab:mmesons1}. 

For tm fermions we have performed
simulations on three volumes and finite volume effects 
turn out to be very small for all the values of the mass.
In the following, we will present only results 
obtained on the $16^3\times32$ lattice. 
For~overlap fermions we extract the pion mass from $C_{P-S}$ 
(Fig.~\ref{fig:mpi_ov_tm} shows how large can be the finite volume effects 
due to the exact zero modes present in $C_P$, 
at small quark masses). The lowest pion mass turns out to be 
very small ($M^{\mathrm{ov}} _\pi\simeq230$ MeV and
$M_{\pi}^{\mathrm{ov}} L_{12}=1.73$) and thus, despite the 
cancellation above, a finite size effect at percent level 
can not be excluded.
In Fig.~\ref{fig:mpi_ov_tm} results for $\mathcal{O}(a)$ improved Wilson 
fermions are also reported. In this case the simulations had to be stopped at 
rather large values of the quark mass to avoid exceptional
configurations. With both, tm and overlap fermions, 
\mbox{we can reach instead very low values of $M_\pi$.}

For overlap fermions, $M^2_\pi$ has, to a very good approximation, 
a linear behaviour with $m_{\rm ov}$ and a linear extrapolation to 
the chiral limit gives an intercept of $-0.002(6)$. 
For tm fermions the behaviour is better described by
a quadratic form and the fit gives an intercept of $0.0054(4)$.
This value, non-compatible with zero, is due to the residual O($a$)
uncertainty in $\kappa_{{\rm cr}}$. This uncertainty also induces  
an O$(a^2\mu^2)$ effect in the pion mass. One can also notice from 
Fig.~\ref{fig:mpi_ov_tm} that the pion masses obtained with tm
always lay above the ones obtained with overlap fermions. 
This is due to the renormalization factor 
$Z_{\rm m}^{{\rm RGI}}$ (needed to obtain the renormalization group
invariant quark mass) that, for tm 
fermions, turns out to be roughly a factor 2.5 larger 
than the corresponding one for overlap fermions (which is close to 
1~\cite{HJLW01}). 

\begin{figure}[t]
\vspace{-0.1cm}
\hspace*{-0.3cm}
\epsfig{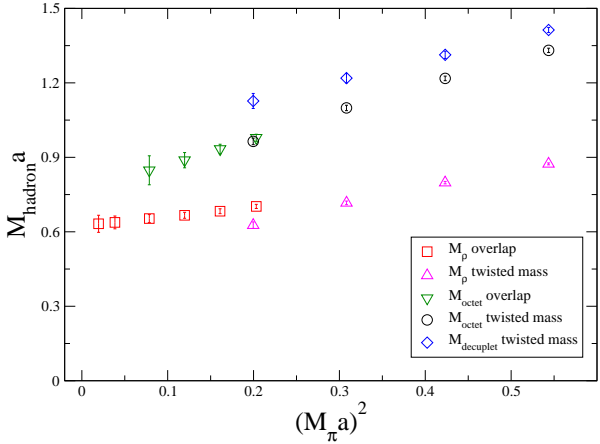}
\vspace{-1.3cm}
\caption{$M_\rho a$ and $M_{\rm baryon} a$ vs. $M_\pi^2 a^2$. 
\label{fig:mhadrons}}
\vspace{-0.7cm}
\end{figure}

The vector meson mass has been extracted from 
$C_{A}^b(x_0)\!=\!\sum_{k=1}^3\sum_{\mathbf x} \langle
A_k^b(x)A_k^b(0)\rangle\;\;(a\!\!=\!\!1,2)$
and $C_{V}^b(x_0) = \sum_{k=1}^3\sum_{\mathbf x}
\langle V_k^b(x)V_k^b(0)\rangle$ in the tm and overlap case respectively.
The results are plotted in 
Fig.~\ref{fig:mhadrons}. In the tm case we observe 
(both with and without smearing) a progressive
worsening of the plateaux for the effective masses as the quark mass
decreases. This phenomenon is particularly evident for the lowest three
masses where, due to these uncertainties, we prefer not to plot any
result. 

We have also computed the proton and the $\Delta^{++}$ correlators with 
interpolating operators
$B_\alpha^{{\rm oct}}=\epsilon^{abc}(d^{a\,T} C\gamma_5
u^b)u^c_\alpha$ and
$B_{k,\alpha}^{{\rm dec}}=\epsilon^{abc}(u^{a\,T} C\gamma_k
u^b)u^c_\alpha$ (with $k=1,2,3$ equivalent) respectively. 
The results are presented in Fig.~\ref{fig:mhadrons}.
In the overlap case, due to the smaller volume, the decuplet
turns out to be too noisy for a reliable estimate.
In the tm case, the same phenomenon as for $M_\rho$ has been 
observed on the lowest three masses.
Notice that, in order to obtain the physical two point correlators,
one has to rotate those computed in the ``twisted'' basis according to
\vspace*{-0.2cm}
\bea
\langle \bar B_\alpha^{{\rm oct,dec}}(x)B_\beta^{{\rm oct,dec}}(0)
\rangle_{{\rm phys}}=
\frac{1}{2}(1+i\gamma_5)_{\alpha\gamma}\nonumber\\
\times\langle\bar B_\gamma^{{\rm oct,dec}}(x)B_\delta^{{\rm oct,dec}}(0)\rangle_{{\rm tb}}(1+i\gamma_5)_{\delta\beta}\;.\nonumber
\eea 
    
\vspace*{-0.1cm}
\subsection{DECAY CONSTANTS}

\begin{figure}[t]
\vspace{-0.0cm}
\hspace*{-0.6cm}
\epsfig{file=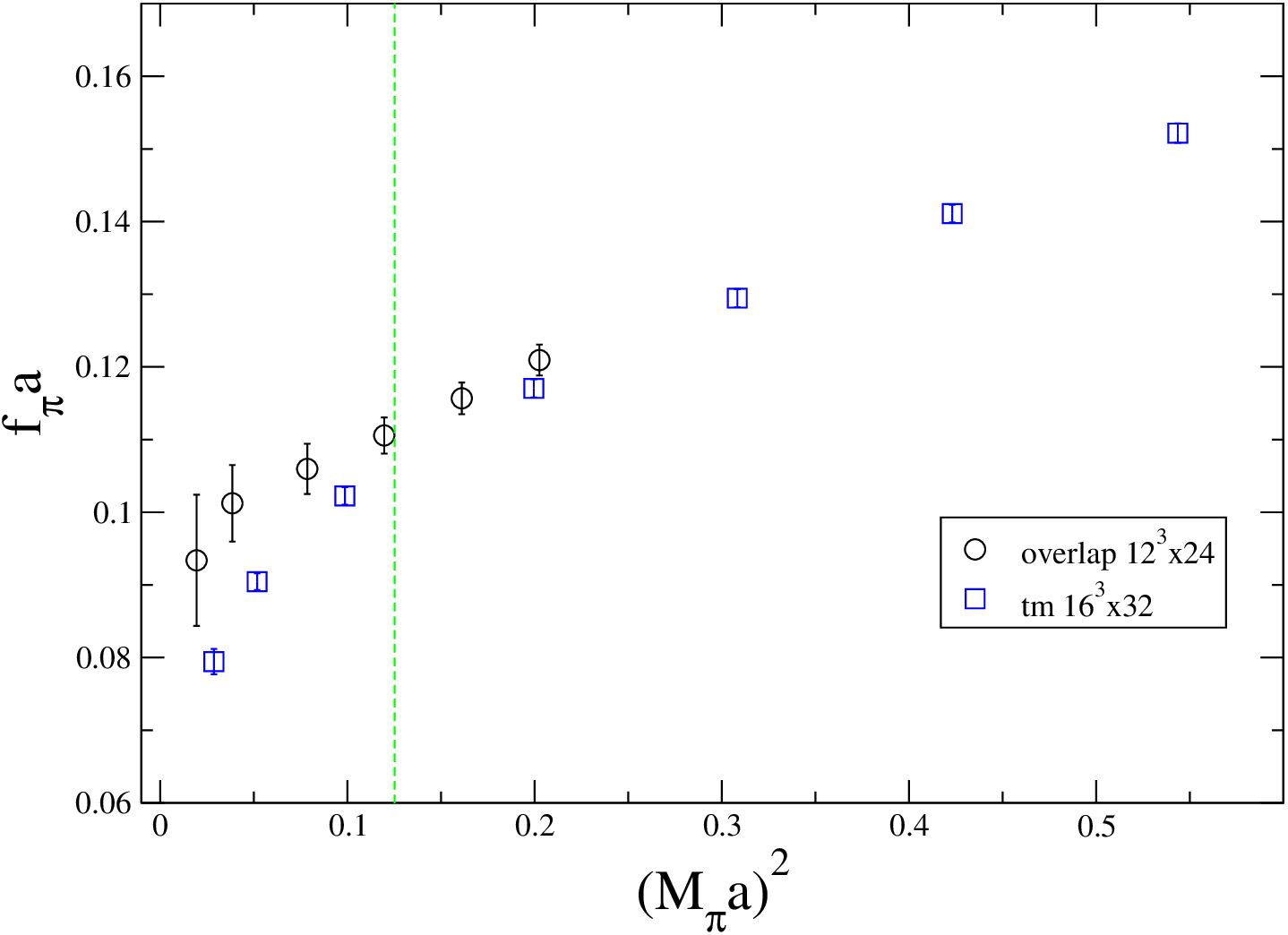,angle=0,width=1.05\linewidth}
\vspace{-0.9cm}
\caption{$f^{\rm ov}_\pi a$ and $f^{\rm tm}_\pi a$ vs. $M_\pi^2 a^2$. 
\label{fig:fpi_ov_tm_final}}
\vspace{-0.75cm}
\end{figure}

\vspace*{+0.1cm}
By using the PCAC  relation, the pseudoscalar decay constants
have been computed (without need of any renormalization
constant) from the ratio  
$f^{{\rm ov}}_\pi=2m_{{\rm bare}}|\langle 0|P|\pi\rangle|/M_\pi^2$,
where $m_{{\rm bare}}$ stands either for $m_{\rm ov}$ or for $\mu$.
Results are reported in Tab.~\ref{tab:mmesons1} and plotted in 
Fig.~\ref{fig:fpi_ov_tm_final}.
At one loop in quenched chiral perturbation theory (qChPT), 
$f_\pi$ has neither chiral logarithms nor finite volume effects: 
$f_\pi=f\left(1+(\alpha_5 M_\pi^2\right/(4 \pi f)^2)$
with $f$ and $\alpha_5$ low energy constants. In the overlap case 
$f_\pi$ nicely 
follows the linear behaviour predicted by qChPT. Neglecting $SU(3)$
breaking effects (which are well below our statistical uncertainty) 
we get $f_\pi=155(11)$ MeV, $f_K=173(8)$ MeV, $f_K/f_\pi=1.11(3)$.

In the case of tm fermions, we observe a bending of 
$f_\pi^\mathrm{tm}$ when the pion mass is small. 
It has been argued~\cite{Frezzotti:2003ni} that for O($a$) improved
quantities, the condition $m_{\rm tm}\gg a^2 \Lambda_{\rm QCD}^3$ 
has to be satisfied in order for the breaking of the chiral symmetry 
not to be driven by the Wilson term. The puzzling fact is that, 
assuming a coefficient of order one, this condition seems to be
satisfied by all of our data points. In Fig.~\ref{fig:fpi_ov_tm_final}, 
the vertical line shows the r.h.s of the stronger condition 
$m_{\rm tm}\gg a \Lambda_{\rm QCD}^2$, which should be valid
for non-improved quantities. We are thus left with the question of which
inequality has to be satisfied: either the weaker one
with a large coefficient or the stronger one with a coefficient of order
one. This phenomenon, together with that observed for $M_\rho$ and
$M_{\rm baryon}$ at the smallest quark masses, requires further 
investigation at smaller values of~$a$. 

Finally we find~\cite{solver} that tm fermions are a factor
of 20-40 faster than overlap fermions and thus
have the potential for dynamical simulations at realistically
small quark masses on the next generation of supercomputers.

\end{document}